\documentstyle[aps,pre,epsfig]{revtex}

\input epsf.sty

\begin{document}
\title{High temperature expansion for a driven bilayer system}
\author{I. Mazilu$^1$ and B. Schmittmann$^2$}
\address{$^1$Department of Physics and Engineering, \\
Washington and Lee University, Lexington, VA 24450;\\
$^2$Center for Stochastic Processes in Science and Engineering,\\
Physics Department, Virginia Tech, Blacksburg, VA 24061-0435, USA.}
\date{January 22, 2003 }
\maketitle

\begin{abstract}
Based directly on the microscopic lattice dynamics, a simple high
temperature expansion can be devised for non-equilibrium steady states. 
We apply this technique to investigate the disordered phase and the phase
diagram for a driven bilayer lattice gas at half filling. Our
approximation captures the phases first observed in simulations,
provides estimates for the transition lines, and allows us to 
compute signature observables of non-equilibrium dynamics, namely, 
particle and energy currents. Its focus on non-universal quantities
offers a useful analytic complement to field-theoretic approaches. 
\end{abstract}

\section{Introduction}

Many-particle systems in a state of thermal equilibrium are the exception,
rather than the rule. Physical reality is overwhelmingly in a
far-from-equilibrium state. Examples range from living cells and weather
patterns to ripples on water and sand. As we leave the framework of standard
Gibbs ensemble theory for equilibrium systems, we have to search for new
avenues and tools, seeking to understand and classify non-equilibrium
behavior. As a first step along this road, the study of the simplest
generalizations of equilibrium systems, i.e., {\em non-equilibrium steady
states} (NESS), has been particularly fruitful \cite{SZ-rev,other-revs}.
Progress has relied predominantly on simulations, mean-field theory and
renormalization group analyses for simple model systems. A class of models
which exhibit especially interesting behavior are driven diffusive systems.
Microscopically, these are lattice gases, consisting of one or more species
of particles and holes, whose densities are conserved. An external driving
force, combined with suitable boundary conditions, maintains a NESS. In the
simplest case \cite{KLS}, a uniform bias, or drive, $E$, is imposed on an
Ising lattice gas such that a nonzero steady-state mass current is induced.
This model differs significantly from the usual Ising model: it displays
generic long-range correlations \cite{KLS,ZWLV,GLMS}, and belongs to a
non-equilibrium universality class \cite{crit} with upper critical dimension 
$d_{c}=5$. The ordered phase is phase-separated into two strips of high vs
low density aligned with the bias. In contrast to equilibrium, bulk and
interfacial properties are inextricably intertwined here \cite{lowT}.

To avoid the complications due to the presence of interfaces, a bilayer
structure was suggested \cite{KKM}: in the two-dimensional case, a second
lattice was introduced, allowing for particle-hole exchanges between each
site and its mirror image. This bilayer system is half filled with
particles, and both layers are driven in the same direction. In the absence
of any energetic couplings between the two layers, it was hoped that typical
ordered configurations would show {\em homogeneous} densities on each layer,
one almost full and the other nearly empty. Remarkably, however, this
expectation proved too naive: Monte Carlo simulations \cite{2l-early} showed
a sequence of{\em \ two} phase transitions, as the temperature is lowered:
the first transition takes the system from a disordered (D) phase to a
strip-like (S) structure showing phase-separation {\em within each layer},
with interfaces parallel to the drive and `on top of' one another. The
anticipated ``full-empty'' (FE) phase, with uniform densities on both
layers, only emerges after a second transition which occurs at a lower
temperature. Once an interaction $J$, of either sign, between nearest
neighbors on different layers is introduced, the full phase diagram in ($J,E$%
) space can be mapped out \cite{HZS,CW}, using Monte Carlo simulations. As
one might expect, the S (FE) phase dominates for attractive (repulsive)
cross-layer coupling $J$. Remarkably, however, there is a small but finite
region where the S-phase prevails even though the cross-layer coupling is
weakly repulsive (cf. Fig. 1). The presence of this domain puts the two
transitions, observed for $J=0$, into perspective. We note for completeness
that universal properties along the lines of continuous transitions have
been analyzed in \cite{TSZ} with the help of renormalized field theory.

To provide additional motivation for the study of layered structures, we
note that multilayer models have a long history in equilibrium statistical
mechanics\cite{ballentine,binder,hansen}. On the theoretical side, they
allow for the study of dimensional crossover \cite{dim-cross}; on the more
applied side, they provide natural models for the analysis of intercalated
systems \cite{intercalation}, interacting solid surfaces or thin films \cite%
{ferrenberg}. Since intercalated systems are often driven by chemical
gradients or electric fields, to speed the diffusion of foreign atoms into
the host material, it is quite natural to study driven layered structures.

Simulations relie, of course, on discrete lattice models. In contrast, field
theories operate in the continuum, and thus, all discrete degrees of freedom
have to be coarse-grained before these powerful techniques can be applied.
In the process, non-universal information is lost, such as, e.g., the
location of transition lines in the phase diagram. It is therefore desirable
to identify a second analytic approach which is based directly on the
microscopic model and thus complements both, simulations and continuum
theories. Fortunately, high temperature expansion techniques \cite{HTS-revs}
can be generalized to interacting driven lattice gases \cite{ZWLV,SZ,LZS}.
For the single-layer case, two-point correlation functions can be computed
approximately \cite{ZWLV,SZ} and display the expected power law decays in
the steady state. With some care, the approximate location of order-disorder
transitions can be extracted and compared to simulation results \cite{SZ}.
Given the nature of the approximation, {\em quantitative} accuracy cannot be
expected, but the {\em qualitative} agreement of data and approximation is
remarkably good.

While the high temperature expansion is quite successful for the usual
driven lattice gas, it is not clear to what extent it is capable of
capturing the main features of other driven systems. This motivates the work
presented in this paper, namely, the analysis of the bilayer system with
this technique. Within a first-order approximation, we compute the two-point
correlation functions and several related quantities, such as the particle
current and the energy flux through the system. We extract the approximate
location of the continuous transition lines and compare our results to the
Monte Carlo data. As in the single-layer case, the qualitative features of
the transition lines are reproduced as well as can be expected. Some
limitations of the method will be discussed.

This paper is organized as follows. We first introduce the bilayer model.\
After a brief summary of the high temperature expansion, we derive the
closed set of equations satisfied by the two-point functions. We then obtain
the solutions and extract the transition lines. Next, we show how the mass
and energy currents through the system can be expressed in terms of pair
correlations. We conclude with some comments and open questions.

\section{The bilayer model}

A variant of the driven Ising model \cite{KLS}, the model consists of two
square lattices, one stacked above the other, resulting in a bilayer
structure of size $L^{2}\times 2$. Each lattice site $\vec{r}\equiv (x,y,z)$%
, with $x,y=1,2,...,L$ and $z=0,1$, carries a spin variable $s(\vec{r})=\pm 1
$. Often, we also use lattice gas language, mapping spins into particles or
holes, via $s(\vec{r})\equiv 2n(\vec{r})-1$. The local occupation variable $%
n(\vec{r})$ takes the values $1$ or $0$, indicating whether a particle is
present or not. The total magnetization, $\sum_{\vec{r}}s(\vec{r})$, is
fixed at zero so that the Ising critical point can be accessed. Within each
layer, nearest-neighbor spins interact through a ferromagnetic exchange
coupling $J_{0}>0$; in contrast, the cross-layer interaction $J$, which
couples spins $s(x,y,0)$ and $s(x,y,1)$, can take both signs. These choices
are motivated by the physics of intercalated systems \cite{intercalation}.
Thus, the Hamiltonian of the system can be written in the form 
\begin{equation}
H=-J_{0}\sum_{z}\sum_{nn}s(x,y,z)s(x^{\prime },y^{\prime
},z)-2J\sum_{x,y}s(x,y,0)s(x,y,1)  \label{H}
\end{equation}%
where $\sum_{nn}$ denotes the sum over all nearest-neighbor pairs $(x,y,z)$
and $(x^{\prime },y^{\prime },z)$ within the same plane. A heat bath at
temperature $T$ is coupled to the system, in order to model thermal
fluctuations. We use fully periodic boundary conditions in all directions;
hence the factor of $2$ in front of the cross-layer coupling $J$.

In the absence of the drive, particles hop to empty nearest-neighbor sites
according to the usual Metropolis \cite{Metropolis} rates, $\min \left\{
1,\exp \left( -\beta \Delta H\right) \right\} $, where $\Delta H$ is the
energy difference due to the jump. Respecting the conservation of density,
the phase diagram of this system is easily found. At high temperatures, a
disordered phase persists, characterized by correlations which fall off
exponentially. At a critical temperature $T_{c}(J)$, a continuous transition
occurs into the S (FE) phase for $J>0$ ($J<0$). At $J=0$, the critical
temperature takes the Onsager value \cite{Onsager} $%
T_{c}(0)=2.269...J_{0}/k_{B}$. For finite $J$, $T_{c}(J)$ is even in $J$,
due to a simple gauge symmetry, and increases monotonically with $\left|
J\right| $. For $J\rightarrow \pm \infty $, nearest-neighbor spin pairs,
with the partners located on different layers, combine into dimers who
couple to neighboring dimers with strength $2J_{0}$. As a result, the
critical temperature approaches the limit $T_{c}(\pm \infty )=2T_{c}(0)$.
The line $J=0$, $T<T_{c}(0)$ is a line of first-order transitions between
the S and FE phases. It ends in a bicritical point at $J=0$, $T=T_{c}(0)$.

To drive the system out of equilibrium, we apply a bias (an ``electric''
field) $\vec{E}$ along the positive $x$-axis. The contents of two sites, $%
\vec{r}$ and $\vec{r}+\hat{a}$, separated by a (unit) lattice vector $ \hat{a}$, are
exchanged according to the rate 
\begin{equation}
c(\vec{r},\vec{r}+\hat{a};\left\{ s\right\} )=\min \left\{ 1,\exp \left[
-\beta \Delta H+\beta \,\hat{a} \cdot \vec{E}\,(n(\vec{r})-n(\vec{r}+\hat{a}))%
\right] \right\}  \label{rates}
\end{equation}
The argument $\left\{ s\right\} $ reminds us that the rate depends on a
local neighborhood of the central pair. Due to $E$, particle hops against
the drive become unfavorable. In conjunction with periodic boundary
conditions in the $x$- and $y$-directions, the system settles into a
non-equilibrium steady state with a net particle current.

The phase diagram, resulting from Monte Carlo simulations at $J_{0}=1$ and
infinite $E$, is shown in Fig. 1. The same phases and transitions are found,
but the bicritical point and its attached first order line are shifted to
higher $T$ and into the $J<0$ region. Thus, the S phase is observed to be
stable in a finite window of negative interlayer coupling, so that two
transitions must occur along the $J=0$ axis. This discovery represents the
most unexpected new characteristic of this driven diffusive system. We also
note the decrease of the critical temperatures for very large $\left|
J\right| $. In a recent paper \cite{CW}, this phase
diagram was extended to include {\em unequal intra-layer%
} attractive couplings. In this case, the bicritical point is shifted
even further into the negative region of $J$ as the coupling transverse to
the bias increases.

We now turn to the analysis of this model in terms of a high temperature
expansion.

\begin{figure}[tbp]
\input{epsf}
\begin{center}
\vspace{-3.8cm}
\hspace{2.5cm}
\begin{minipage}{0.8\textwidth}
  \epsfxsize = 0.7\textwidth \epsfysize = 0.7\textwidth 
  \epsfbox{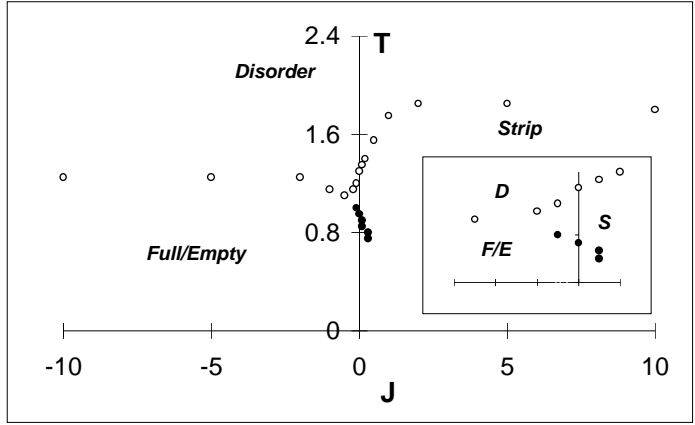}
    \vspace{-0.5cm}
\end{minipage}
\end{center}
\caption{The phase diagram of the driven bilayer lattice gas from Monte
Carlo simulations \protect\cite{HZS}, at $J_{0}=1$ and
infinite $E$. $T$ is measured in units of the
Onsager temperature, $T_{c}(0)$.  The domains of the three phases are
indicated. The inset demonstrates the shift of the bicritical point to
negative $J$. Open (filled) circles indicate continuous (first order) 
transitions.}
\end{figure}


\section{High temperature expansion}

The dynamics underlying the\ Monte Carlo simulations is easily expressed via
a master equation. The latter provides a convenient starting point for a
high temperature expansion. For simplicity, we take the thermodynamic limit
within each plane, i.e., $L\rightarrow \infty $. Following \cite{ZWLV}, we
first derive the equations of motion for the two-point functions. By virtue
of the familiar hierarchy, they are coupled to the three-point functions;
however, we will argue that these are negligible (while non-zero, they are
numerically rather small), so as to arrive at a closed system of equations
for the two-point correlations. Temperature appears in these equations
through the rates, via the combinations $\beta J$, $\beta J_{0}$, and $\beta
E$. To preserve the non-equilibrium nature of our dynamics, we expand in $%
\beta J$ and $\beta J_{0}$, keeping $\beta E$ finite. Technically, this
requires that $E$ always dominates the energetic contribution, i.e., $%
E>\Delta H$ for all jumps along $E$. To first order, a linear, ${\em %
inhomogeneous}$ system of equations results, which can be solved exactly %
\cite{SZ} and forms the basis of our analysis.

\subsection{The equations of motion and their solution.}

Before turning to any detailed calculations, let us introduce the key
quantities. The {\em two-point correlation function} is defined as: 
\begin{equation}
G(\vec{r}-\vec{r}\text{ }^{\prime })=\left\langle s(\vec{r})s(\vec{r}\text{ }%
^{\prime })\right\rangle  \label{CF}
\end{equation}
where $\left\langle \cdot \right\rangle $ denotes the configurational
average. Due to translation invariance, $G$ depends only on the difference
of the two vectors. Moreover, $G$ is invariant under reflection of one or
several lattice directions; e.g., $G(x,y,z)=G(-x,y,z)$, etc. The correlation
function at the origin is obviously unity, $G(\vec{0})=\left\langle s^{2}(%
\vec{r})\right\rangle =1$. We also introduce the Fourier transform of $G$,
i.e., the{\bf \ }{\em structure factor}: 
\begin{equation}
S(k,p,q)\equiv \sum_{z=0,1}\sum_{x,y=-\infty }^{\infty
}G(x,y,z)e^{-i(kx+py+qz)}  \label{SF}
\end{equation}
Since we take the thermodynamic limit $L\rightarrow \infty $, the wave
vectors $k$ and $p$ are continuous, but restricted to the first Brillouin
zone $[-\pi ,\pi ]$, while $q$ is discrete, taking only the two values $0$
and $\pi $. For completeness, we also give the inverse transform, 
\begin{eqnarray}
G(x,y,z) &=&\frac{1}{2(2\pi )^{2}}\sum_{q=0,\pi }\int_{-\pi }^{+\pi
}dk\int_{-\pi }^{+\pi }dpS(k,p,q)e^{i(kx+py+qz)}  \nonumber \\
&\equiv &\int S(k,p,q)e^{i(kx+py+qz)}  \label{FT}
\end{eqnarray}
where the second line just defines some simplified notation.

To set up the high temperature expansion, we first define the actual
expansion parameters of our theory, namely 
\begin{eqnarray}
K_{0} &\equiv &\beta J_{0}  \nonumber \\
K &\equiv &\beta J  \label{Ks}
\end{eqnarray}%
For $K=K_{0}=0$, the steady-state distribution is exactly known \cite%
{spitzer} to be uniform for all $E$: $P^{\ast }\propto 1$, so that we are
expanding about a well-defined zeroth order solution. The correlation
functions and structure factors for this limit are trivial, namely, $G(\vec{r%
})=\delta _{\vec{r},\vec{0}}$ where $\delta $ denotes the Kronecker symbol,
and $S(k,p,q)=1$. Returning to the interacting case, we note that $G(\vec{r})
$, for $\vec{r}\neq \vec{0}$, is already of first order in the small
parameter. Similarly, we can write the structure factor as a sum of two
terms. The first term is just the zeroth order solution, while the second, $%
\tilde{S}$, carries the information about the interactions, 
\begin{equation}
S(k,p,q)=1+\tilde{S}(k,p,q)  \label{S}
\end{equation}%
so that we can recast $G(\vec{r})$, for $\vec{r}\neq \vec{0}$, in the form 
\begin{equation}
G(x,y,z)=\int \widetilde{S}(k,p,q)e^{i(kx+py+qz)}\text{\ for }x,y,z\neq 0
\label{G}
\end{equation}

The exact equations of motion for $G$ are easily derived from the master
equation \cite{ZWLV}: 
\begin{equation}
{\displaystyle{d \over dt}}%
\left\langle s(\vec{r})s(\vec{r}\text{ }^{\prime })\right\rangle =\sum_{\vec{%
x},\vec{x}^{\prime }}\left\langle s(\vec{r})s(\vec{r}\text{ }^{\prime })%
\left[ s(\vec{x})s(\vec{x}\text{ }^{\prime })-1\right] c\left( \vec{x},\vec{x%
}\text{ }^{\prime };\left\{ s\right\} \right) \right\rangle  \label{eom}
\end{equation}
Here, the sum runs over{\em \ nearest-neighbor} pairs ($\vec{x},\vec{x}$ $%
^{\prime }$) such that $\vec{x}\in \left\{ \vec{r},\vec{r}\text{ }^{\prime
}\right\} $ but $\vec{x}$ $^{\prime }\notin \left\{ \vec{r},\vec{r}\text{ }%
^{\prime }\right\} $. Stationary correlations are obtained by setting the
left hand side to zero. Clearly, jumps along and against all three lattice
directions will contribute to the right hand side of Eq.~(\ref{eom}).

To proceed, let us write the jump rates in a form which makes their
dependence on the spin configuration $\left\{ s\right\} $ explicit, so that
the configurational averages in Eq.~(\ref{eom}) can be performed. For {\em %
infinite} drive, a particle jumps along the field with rate unity, but never
against it, so that the transition rates {\em parallel to the field} can be
written as: 
\begin{equation}
c_{\Vert }^{\infty }\left( \vec{r},\vec{r}+\hat{x};\left\{ s\right\} \right)
=%
{\displaystyle{1 \over 4}}%
\left[ s(\vec{r})-s(\vec{r}+\hat{x})+2\right]  \label{c_par_Einf}
\end{equation}
Here, $\hat{x}$ is a unit vector in the positive $x$-direction.
In the case of {\em finite} drive, our restriction $E>\Delta H$ ensures that
jumps along $E$ still occur with unit rate, while those against $E$ are
suppressed by a factor of $\exp \left[ -\beta \left( \Delta H+E\right) %
\right] $. Defining 
\begin{equation}
\varepsilon \equiv e^{-\beta E}  \label{eps}
\end{equation}
Eq.~(\ref{c_par_Einf}) must be amended to 
\begin{equation}
c_{\Vert }\left( \vec{r},\vec{r}+\hat{x};\left\{ s\right\} \right) =%
{\displaystyle{1 \over 4}}%
\left[ s(\vec{r})-s(\vec{r}+\hat{x})+2\right] +%
{\displaystyle{\varepsilon  \over 4}}%
\left[ s(\vec{r}+\hat{x})-s(\vec{r})+2\right] \exp (-\beta \Delta H)
\label{c_par}
\end{equation}
Transverse to the field we have two jump rates, corresponding to the two
transverse directions ($y$ and $z$). Both of these are regulated by the
energy difference due to a jump: 
\begin{equation}
c_{\bot }\left( \vec{r},\vec{r}+\hat{a};\left\{ s\right\} \right) =\min
\left\{ 1,\exp \left( -\beta \Delta H\right) \right\}  \label{c_perp}
\end{equation}
We are now ready to expand the rates in powers of $K$ and $K_{0}$ while
keeping $\varepsilon $ finite:\smallskip 
\begin{eqnarray}
c_{\Vert }\left( \vec{r},\vec{r}+\hat{x};\left\{ s\right\} \right) &=&%
{\displaystyle{1 \over 4}}%
\left[ s(\vec{r})-s(\vec{r}+\hat{x})+2\right] +%
{\displaystyle{\varepsilon  \over 4}}%
\left[ s(\vec{r}+\hat{x})-s(\vec{r})+2\right] \left( 1-\beta \Delta H\right)
+O(\beta ^{2})  \label{c_par_exp} \\
c_{\bot }\left( \vec{r},\vec{r}+\hat{a};\left\{ s\right\} \right) &=&1+\beta
c_{2}\left( \vec{r},\vec{r}+\hat{a};\left\{ s\right\} \right) +O(\beta ^{2})
\label{c_perp_exp}
\end{eqnarray}
with 
\begin{equation}
c_{2}\left( \vec{r},\vec{r}+\hat{a};\left\{ s\right\} \right) =-\frac{1}{2}%
(\Delta H+\left| \Delta H\right| )
\end{equation}

Given these simple forms for the rates, we can now derive the equations of
motion satisfied by the pair correlations directly from Eq.~(\ref{eom}),
following \cite{ZWLV}. A few details are outlined in the Appendix. Keeping
only corrections to first order in $K$, $K_{0}$ and neglecting three-point
correlations, we obtain a {\em closed set of linear equations} for $G(x,y,z)$%
.{\bf \ }Since the dynamics is restricted to nearest-neighbor processes, it
is not surprising that the equations involve an anisotropic lattice
Laplacian acting on $G(x,y,z)$. For $x,y,z$ near the origin, the Laplacian
may include the origin and will thus generate inhomogeneities in the system
of equations. The detailed form depends on the chosen boundary conditions,
and, of course, on the three parameters $K$, $K_{0}$, and $\varepsilon $.
Below, we show the set of equations for fully periodic boundary conditions.
The first three equations result from nearest neighbors of the origin, $\vec{%
r}=(1,0,0)$, $(0,1,0)$, and $(0,0,1)$: 
\begin{eqnarray}
\partial _{t}G(1,0,0) &=&(1+\varepsilon
)[G(2,0,0)-G(1,0,0)]+4[G(1,1,0)-G(1,0,0)]  \nonumber \\
&&+4[G(1,0,1)-G(1,0,0)]+2\varepsilon K_{0}+8K_{0}  \nonumber \\
\partial _{t}G(0,1,0) &=&2(1+\varepsilon
)[G(1,1,0)-G(0,1,0)]+2[G(0,2,0)-G(0,1,0)]  \nonumber \\
&&+4[G(0,1,1)-G(0,1,0)]+4\varepsilon K_{0}+6K_{0}  \label{G-10} \\
\partial _{t}G(0,0,1) &=&2(1+\varepsilon
)[G(1,0,1)-G(0,0,1)]+4[G(0,1,1)-G(0,0,1)]  \nonumber \\
&&+8K+8\varepsilon K  \nonumber
\end{eqnarray}
\ \ \ By virtue of invariance under reflections, these equations also hold
for the other nearest neighbors $\vec{r}=(-1,0,0)$, $(0,-1,0)$, and $%
(0,0,-1) $. The following three equations arise from the next-nearest
neighbor sites, $\vec{r}=(1,1,0)$, $(0,1,1)$, and $(1,0,1)$, and their
reflections: 
\begin{eqnarray}
\partial _{t}G(1,1,0) &=&(1+\varepsilon
)[G(2,1,0)+G(0,1,0)-2G(1,1,0)]+2[G(1,2,0)+G(1,0,0)  \nonumber \\
&&-2G(1,1,0)]+\newline
4[G(1,1,1)-G(1,1,0)]-2K_{0}-2\varepsilon K_{0}\newline
\nonumber \\
\partial _{t}G(0,1,1) &=&2(1+\varepsilon
)[G(1,1,1)-G(0,1,1)]+2[G(0,2,1)+G(0,0,1)-2G(0,1,1)]  \nonumber \\
&&+\newline
4[G(0,1,0)-G(0,1,1)]-4[K_{0}+K]\newline
\label{G-11} \\
\partial _{t}G(1,0,1) &=&(1+\varepsilon
)[G(2,0,1)+G(0,0,1)-2G(1,0,1)]+4[G(1,1,1)-G(1,0,1)]  \nonumber \\
&&+\text{\newline
}4[G(1,0,0)-G(1,0,1)]-4\varepsilon K-4K_{0}  \nonumber
\end{eqnarray}
Increasing the separation of the participating sites further, to $\vec{r}%
=(2,0,0)$ and $(0,2,0)$, we obtain: 
\begin{eqnarray}
\partial _{t}G(2,0,0) &=&(1+\varepsilon
)[G(3,0,0)+G(1,0,0)-2G(2,0,0)]+4[G(2,1,0)-G(2,0,0)]  \nonumber \\
&&+4[G(2,0,1)-G(2,0,0)]-2\varepsilon K_{0}  \nonumber \\
\partial _{t}G(0,2,0) &=&2(1+\varepsilon
)[G(1,2,0)-G(0,2,0)]+2[G(0,3,0)+G(0,1,0)-2G(0,2,0)]  \label{G-20} \\
&&+4[G(0,2,1)-G(0,2,0)]\newline
-2K_{0}  \nonumber
\end{eqnarray}
And finally, all $G$'s with $\left| x\right| +\left| y\right| +\left|
z\right| >2$ satisfy homogeneous equations: 
\begin{eqnarray}
\partial _{t}G(i,j,k) &=&(1+\varepsilon )[G(i+1,j,k)+G(i-1,j,k)-2G(i,j,k)] 
\nonumber \\
&&+2[G(i,j+1,k)+G(i,j-1,k)-2G(i,j,k)]+\newline
4[G(i,j,k-1)-G(i,j,k)]  \label{G-rest}
\end{eqnarray}
The last equation contains the full anisotropic lattice Laplacian, acting on 
$G(i,j,k)$, without any inhomogeneities being generated. We note, for
further reference, that the right hand sides of Eqns (\ref{G-10}-\ref{G-rest}%
) contain contributions from exchanges along and against the three lattice
directions. Starting from Eq.~(\ref{eom}), it is of course easy to keep
track of terms originating in transverse vs parallel jumps. Below, this
distinction will become important when we turn to energy currents.

To solve this system, we closely follow the method presented in \cite{SZ}.
Returning to Eq.~(\ref{S}), we need to focus only on $\tilde{S}$, since this
quantity carries the information about the interactions. Recalling Eq.~(\ref%
{G}), we first express $G$ through its Fourier transform $\tilde{S}$,
exploiting translation invariance and linearity. Then, we invoke the
completeness of complex exponentials to project out an equivalent set of
(algebraic) equations for $\tilde{S}$.

To follow through with this program, we first define the anisotropic lattice
Laplacian in Fourier space, 
\begin{equation}
\delta (k,p,q)\equiv 2(1+\varepsilon )(1-\cos k)+4(1-\cos p)+4(1-\cos q)
\label{Lap}
\end{equation}%
and second, introduce the three (as yet unknown) quantities: 
\begin{eqnarray}
I_{1} &\equiv &\int \tilde{S}(1-\cos k)  \nonumber \\
I_{2} &\equiv &\int \tilde{S}(1-\cos p)  \label{Is} \\
I_{3} &\equiv &\int \tilde{S}(1-\cos q)  \nonumber
\end{eqnarray}%
With these definitions, the system can be expressed in terms of $\tilde{S}$,
resulting in: 
\begin{eqnarray}
2\varepsilon K_{0}+8K_{0} &=&\int \tilde{S}\delta \exp (ik)+(1+\varepsilon
)I_{1}  \nonumber \\
4\varepsilon K_{0}+6K_{0} &=&\int \tilde{S}\delta \exp (ip)+2I_{2}  \nonumber
\\
8\varepsilon K\newline
+8K &=&\int \tilde{S}\delta \exp (iq)+4I_{3}  \nonumber \\
-2\varepsilon K_{0}-2K_{0} &=&\int \tilde{S}\delta \exp (i(k+p))  \nonumber
\\
-4(K_{0}+K) &=&\int \tilde{S}\delta \exp (i(p+q)) \\
-4\varepsilon K-4K_{0} &=&\int \tilde{S}\delta \exp (i(k+q))  \nonumber \\
-2\varepsilon K_{0} &=&\int \tilde{S}\delta \exp (2ik)  \nonumber \\
-2K_{0} &=&\int \tilde{S}\delta \exp (2ip)  \nonumber \\
0 &=&\int \tilde{S}\delta \exp (i(kx+py+qz))\text{ for }\left| x\right|
+\left| y\right| +\left| z\right| >2  \nonumber
\end{eqnarray}%
To proceed, we treat $I_{1},I_{2},I_{3}$ for the time being as simple
coefficients and move them to the left-hand side. Finally, we need one
additional equation for $x=y=z=0$, which is easily obtained: 
\[
\int \tilde{S}\delta =\int \tilde{S}[2(1+\varepsilon )(1-\cos k)+4(1-\cos
p)+4(1-\cos q)]=2(1+\varepsilon )I_{1}+4I_{2}+4I_{3}
\]%
Now, we are ready to invoke the completeness relation for complex
exponentials, namely, 
\begin{equation}
\sum_{x,y,z}\exp [i(kx+py+qz)]=2(2\pi )^{2}\delta (k)\delta (p)\delta _{q,0}
\end{equation}%
which allows us to solve for $\tilde{S}$: 
\begin{equation}
\tilde{S}(k,p,q)=%
{\displaystyle{L(k,p,q) \over \delta (k,p,q)}}
\label{S-tilde}
\end{equation}%
where 
\begin{eqnarray}
L(k,p,q) &\equiv &2(1+\varepsilon )\left( 1-\cos k\right) I_{1}+4\left(
1-\cos p\right) I_{2}+4\left( 1-\cos q\right) I_{3}  \nonumber \\
&&+\left( 2\varepsilon K_{0}+8K_{0}\right) 2\cos k+\left( 4\varepsilon
K_{0}+6K_{0}\right) 2\cos p  \nonumber \\
&&+\left( 8\varepsilon K\newline
+8K\right) \cos q-\left( 2\varepsilon K_{0}+2K_{0}\right) 4\cos k\cos p
\label{L} \\
&&-\left( 4\varepsilon K+4K_{0}\right) 2\cos k\cos q-4(K_{0}+K)2\cos p\cos q
\nonumber \\
&&-4\varepsilon K_{0}\cos 2k-4K_{0}\cos 2p  \nonumber
\end{eqnarray}%
However, Eq.~(\ref{S-tilde}) is not yet a fully explicit solution for $%
\tilde{S}$, due to the appearance of the three integrals $I_{1}$, $I_{2}$,
and $I_{3}$ in $L$. To determine these three coefficients, we need three
linearly independent equations. One of these equations is given by the value
of $G$ at the origin, $1=G(0,0,0)=\int (1+\tilde{S})$, and the remaining two
can be obtained directly from the definitions of $I_{1}$ and $I_{3}$ in Eq.~(%
\ref{Is}): 
\begin{eqnarray}
0 &=&\int 
{\displaystyle{L(k,p,q) \over \delta (k,p,q)}}
\nonumber \\
0 &=&-I_{1}+\int 
{\displaystyle{L(k,p,q) \over \delta (k,p,q)}}%
(1-\cos k)  \label{matrix} \\
0 &=&-I_{3}+\int 
{\displaystyle{L(k,p,q) \over \delta (k,p,q)}}%
(1-\cos q)  \nonumber
\end{eqnarray}%
After inserting Eq.~(\ref{L}) for $L$, this leads to a set of three
inhomogeneous, linear equations for the three unknowns $I_{1}$, $I_{2}$, and 
$I_{3}$, which are easily solved. Since the details of the associated matrix
inversion are straightforward but tedious, we relegate a few details to the
Appendix. We just note the following overall features: ({\em i}) All three
coefficients are functions of $K$, $K_{0}$, and $\varepsilon $; ({\em ii})
for the whole range of fields $\varepsilon $ and for $K_{0}=1$ and $K=\pm 1$
(attractive and repulsive inter-layer interactions), $I_{1}$ and $I_{2}$ are
negative, while $I_{3}$ is positive for $K=-1$ and negative for $K=+1$.

This concludes the calculation of the structure factor. To summarize, we
find 
\begin{equation}
S(k,p,q)=1+%
{\displaystyle{L(k,p,q) \over \delta (k,p,q)}}%
+O(K^{2},K_{0}^{2},KK_{0})  \label{S-final}
\end{equation}
Even at the lowest nontrivial order, this solution carries a significant
amount of information about the phase diagram of our system. In particular,
we can extract an approximate shape of the critical lines, as we will show
in the following.

\subsection{The critical lines.}

The location of a continuous phase transition is marked by the divergence of
a suitably chosen structure factor, as a function of the external control
parameters. For example, we can locate the order-disorder phase transition
of the usual, two-dimensional Ising model by seeking those values of
temperature (at zero magnetic field) for which the structure factor, $S(\vec{%
k})$, diverges. In the absence of a conservation law, the only singularity
occurs at the Onsager temperature if $\vec{k}=0$, indicating that the system
orders into a spatially homogeneous state. For a lattice gas, however, $S(%
\vec{0})$ is fixed by the conservation law, and we need to seek the onset of
phase {\em separation}, i.e., the emergence of macroscopic spatial
inhomogeneities in the system. In this case, singular behavior occurs in $%
\lim_{\vec{k}\rightarrow 0}S(\vec{k})$, provided the system is half-filled
and tuned to the Onsager temperature.

For the bilayer system, we need to locate, and distinguish, {\em two types}
of continuous transitions, namely, from disorder (D)\ into the strip (S) and
the full-empty (FE) phases, respectively. Since the D-S transition is marked
by the appearance of phase-separated strips in each layer, aligned with the
driving force, it can be located by seeking singularities in $%
\lim_{p\rightarrow 0}S(0,p,0)$. In contrast, the D-FE transition exhibits
homogeneous, but opposite magnetizations in the two planes, so that it can
be found by considering $S(0,0,\pi )$. In fact, these two structure factors
were precisely the order parameters chosen in the MC studies \cite{HZS}.

Yet, another subtlety must be considered: in a typical high temperature
expansion such as ours, only a finite number of terms can be computed.
Hence, any perturbative result for the structure factor must be finite, and
instead, the radius of convergence of the expansion must be estimated. Even
this is not practical here, since we have only two terms of the series. To
circumvent these restrictions \cite{SZ}, we extract the singularity by
looking for {\em zeros} of $S^{-1}$, to first order in $K$ and $K_{0}$.

Starting from Eq.~(\ref{S-final}), we obtain 
\begin{equation}
S^{-1}(k,p,q)=1-%
{\displaystyle{L(k,p,q) \over \delta (k,p,q)}}%
+O(K^{2},K_{0}^{2},KK_{0})  \label{S-1}
\end{equation}%
and seek to locate the zeros of $\lim_{p\rightarrow 0}S^{-1}(0,p,0)$ for the
D-S transition, and of $S^{-1}(0,0,\pi )$ for the D-FE transition. Of
course, we should ensure that these are the first zeros which are
encountered upon lowering the temperature. Therefore, we consider, more
generally, the behavior of $S^{-1}(k,p,q)$ at small $k,p$ and fixed $q$. The
denominator of Eq.~(\ref{S-1}), being the lattice Laplacian, is positive
definite: 
\begin{equation}
\lim_{k,p\rightarrow 0}\delta (k,p,q)=\ (1+\varepsilon
)k^{2}+2p^{2}+4(1-\cos q)+O(k^{4},p^{4},k^{2}p^{2})
\end{equation}%
and vanishes only at the origin. Similarly, we obtain 
\begin{eqnarray}
\lim_{k,p\rightarrow 0}L(k,p,q) &=&16K_{0}\left( 1-\cos q\right) +4\left(
1-\cos q\right) I_{3} \\
&&+k^{2}\left[ (1+\varepsilon )I_{1}+10\varepsilon K_{0}-4K_{0}+\left(
4\varepsilon K+4K_{0}\right) \cos q\right]  \nonumber \\
&&+p^{2}\left[ 2I_{2}+6K_{0}+(4K_{0}+4K)\cos q\right]
+O(k^{4},p^{4},k^{2}p^{2})  \nonumber
\end{eqnarray}%
We note, briefly, that the anisotropic momentum dependence of numerator and
denominator leads to power law correlations in the $x$- and $y$-directions %
\cite{ZWLV,SZ,AGMA}. Combining our results so far, it is apparent that the
zeros of $S^{-1}$ are identical to those of $\delta -L$ in Eq.~(\ref{S-1}).
To simplify notation, we write 
\begin{equation}
\lim_{k,p\rightarrow 0}\left[ \delta (k,p,q)-L(k,p,q)\right] \equiv \tau
_{\Vert }(q)k^{2}+2\tau _{\bot }(q)p^{2}+4\tau _{z}(1-\cos
q)+O(k^{4},p^{4},k^{2}p^{2})
\end{equation}%
and read off 
\begin{eqnarray}
\tau _{\Vert }(q) &=&(1+\varepsilon )\left( 1-I_{1}\right) -10\varepsilon
K_{0}+4K_{0}-\left( 4\varepsilon K+4K_{0}\right) \cos q  \nonumber \\
\tau _{\bot }(q) &=&1-I_{2}-3K_{0}-(2K_{0}+2K)\cos q  \label{tau} \\
\tau _{z} &=&1-I_{3}-4K_{0}  \nonumber
\end{eqnarray}%
In a field-theoretic context \cite{TSZ}, these quantities play the role of
diffusion coefficients: $\tau _{\Vert }$ and $\tau _{\bot }$ control the
in-plane diffusion in the parallel and transverse directions, respectively,
while $\tau _{z}$ controls the cross-plane hopping.

For high temperatures, i.e., small values of $K_{0}=\beta J_{0}$ and $%
K=\beta J$, all three $\tau $-coefficients are positive. Seeking zeros of
these expressions, as $K_{0}$ and $K$ increase, we need to consider the two
cases $q=0$ and $q=\pi $ separately. For $q=0$, we find that $\tau _{\bot
}(0)$ has a single zero at a critical $\beta _{c}^{S}$, for given $J_{0}$, $%
J $ and $\varepsilon $. At these parameter values, $\tau _{\Vert }(0)$ and $%
\tau _{z}$ remain positive. Similarly, for $q=\pi $, the coefficient $\tau
_{z}$ is the one which vanishes first as $\beta $ increases, reaching zero
at a critical $\beta _{c}^{FE}$. Converting into temperatures, we obtain two
functions, $T_{c}^{S}(J_{0},J,\varepsilon )$ and $T_{c}^{FE}(J_{0},J,%
\varepsilon )$, and we need to identify the larger of the two:\ If $\max %
\left[ T_{c}^{S}(J_{0},J,\varepsilon ),T_{c}^{FE}(J_{0},J,\varepsilon )%
\right] =T_{c}^{S}(J_{0},J,\varepsilon )$, the order-disorder transition is
of the D-S type. Otherwise, if $\max \left[ T_{c}^{S}(J_{0},J,\varepsilon
),T_{c}^{FE}(J_{0},J,\varepsilon )\right] =T_{c}^{FE}(J_{0},J,\varepsilon )$%
, the FE phase is selected upon crossing criticality.

While the two critical lines, $T_{c}^{S}(J_{0},J,\varepsilon )$ and $%
T_{c}^{FE}(J_{0},J,\varepsilon )$, can in principle be found analytically,
the details are not particularly illuminating. Instead, we present a range
of numerical results below. For example, for infinite $E$ ($\varepsilon =0$%
), we obtain 
\begin{eqnarray}
k_{B}T_{c}^{S}(J_{0},J,0) &=&4.39J_{0}+2.11J  \label{eps=0} \\
k_{B}T_{c}^{FE}(J_{0},J,0) &=&4.14J_{0}\ -1.36J  \nonumber
\end{eqnarray}%
For finite $E$ with $\varepsilon =\exp (-\beta E)=0.5$, all coefficients
decrease and we find 
\begin{eqnarray}
k_{B}T_{c}^{S}(J_{0},J,0.5) &=&4.15J_{0}+2.03J  \label{eps=0.5} \\
k_{B}T_{c}^{FE}(J_{0},J,0.5) &=&4.05J_{0}\ -1.70J  \nonumber
\end{eqnarray}%
In each case, the bicritical point is defined through the solution of $%
T_{c}^{S}(J_{0},J,\varepsilon )=T_{c}^{FE}(J_{0},J,\varepsilon )$.\ For
comparison, we also quote the equilibrium ($E=0$) results (see the Appendix
for details): 
\begin{eqnarray}
k_{B}T_{c}^{S}(J_{0},J,1) &=&4J_{0}+2J  \label{equ} \\
k_{B}T_{c}^{FE}(J_{0},J,1) &=&4J_{0}-2J  \nonumber
\end{eqnarray}%
which exhibit the expected $J\rightarrow -J$ symmetry.

Recalling that the MC simulations were performed at fixed, positive in-plane
coupling $J_{0}$, we need to consider only the dependence on the cross-plane
coupling $J$ which may take either sign. All of our results show that, for
positive $J$, the D-S transition dominates while, for {\em sufficiently
negative} $J$, a D-FE transition is found. In the following, we discuss the
non-equilibrium ($E\neq 0$) results in more detail.

Fig. 2 shows the critical lines for two typical values of the parameters, $%
\varepsilon =0.5$ and $J_{0}=1$. Being the result of a first order
approximation, the critical lines must of course be linear in $J$.
Therefore, quantitative agreement with the simulation data cannot be
expected; nevertheless, several important features are reproduced: {\em the
existence of two ordered phases }and the{\em \ shift of the bicritical point}
to higher values of $T$ and negative $J$. As a result, the S phase survives
for small, negative $J$, despite being energetically unfavorable. This
phenomenon can be explained qualitatively \cite{HZS} by noting that {\em %
long-range negative} correlations transverse to $E$ dominate the ordering
process for positive $J$, and this mechanism continues to be effective for a
small region of negative $J$. For large and negative $J$, the disordered
state orders into the full-empty (FE) phase, characterized by the planes
being mainly full or empty.
Finally, we comment on the dependence of the critical lines, specifically $%
T_{c}^{S}(1,1,\varepsilon )$ and $T_{c}^{FE}(1,-1,\varepsilon )$, on the
field parameter $\varepsilon =\exp \left( -\beta E\right) $, shown in Fig.
3. For $E=0$, both temperatures are equal, by virtue of the $J\rightarrow -J$
symmetry of the equilibrium system. As the field becomes stronger, the
critical temperature of the D-S transition increases, in contrast to the
critical temperature of the D-FE \ transition which decreases. This behavior
agrees qualitatively with the trend observed in the simulations \cite{HZS,CH}.

\begin{figure}[tbp]
\input{epsf}
\begin{center}
\vspace{-2.5cm}
\hspace{2.5cm}
\begin{minipage}{0.7\textwidth}
  \epsfxsize = 0.7\textwidth \epsfysize = 0.7\textwidth 
  \epsfbox{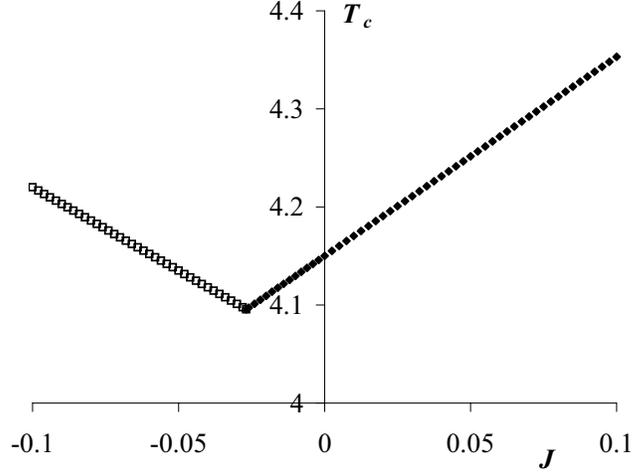}
\end{minipage}
\end{center}
\caption{The dependence of the critical temperature, $\max \left[
T_{c}^{S}(J_{0},J,\protect\varepsilon ),T_{c}^{FE}(J_{0},J,\protect%
\varepsilon )\right] $, on the cross-layer coupling $J$, for $\protect%
\varepsilon =0.5$ and $J_{0}=1$. Filled diamonds (open squares) indicate
that the transition is of the D-S (D-FE) type. Critical temperatures for $%
J>0.1$ ($J<-0.1$) can be obtained by linear extrapolation of the D-S (D-FE)
branch.}
\end{figure}

There are several other quantities of physical interest which are
immediately related to the two-point correlations, such as the steady-state
particle and energy currents. To extract these, we first discuss the inverse
Fourier transform of the structure factor, focusing specifically on the
nearest-neighbor correlations.

\begin{figure}[tbp]
\input{epsf}
\begin{center}
\vspace{-2.5cm}
\hspace{2.5cm}
\begin{minipage}{0.7\textwidth}
  \epsfxsize = 0.7\textwidth \epsfysize = 0.7\textwidth 
  \epsfbox{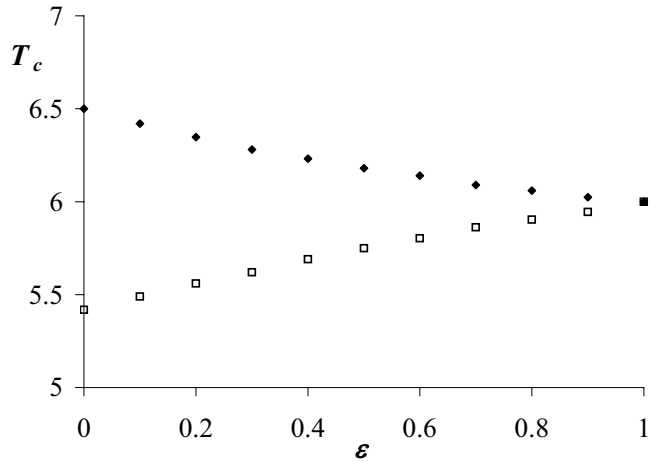}
\end{minipage}
\end{center}
\caption{The critical temperatures, $T_{c}^{S}(1,1,\protect\varepsilon )$
corresponding to a D-S transition (filled diamonds) and $T_{c}^{FE}(1,-1,%
\protect\varepsilon )$ corresponding to a D-FE transition (open squares) as
functions of the field parameter $\protect\varepsilon =\exp (-\protect\beta %
E)$.}
\end{figure}

\subsection{Related physical observables.}

{\em Nearest-neighbor correlations. }These are easily found from our
solution for the structure factor, Eq.~(\ref{S-final}). For example, the
nearest-neighbor correlation in the field direction is given by: 
\begin{equation}
G(1,0,0)=\int \tilde{S}(k,p,q)\cos k+O(K^{2},K_{0}^{2},KK_{0})
\end{equation}%
Since $\int \tilde{S}=0$ by virtue of $G(0,0,0)=1$, we obtain 
\begin{equation}
G(1,0,0)=-\int \tilde{S}(k,p,q)\left( 1-\cos k\right)
+O(K^{2},K_{0}^{2},KK_{0})=-I_{1}
\end{equation}%
and similarly, 
\begin{eqnarray}
G(0,1,0) &=&-I_{2} \\
G(0,0,1) &=&-I_{3}  \nonumber
\end{eqnarray}%
These three integrals are already known since they were required for the
discussion of the critical lines. Specifically, for $\varepsilon =0.5$ we
find, neglecting corrections of $O(K^{2},K_{0}^{2},KK_{0})$:%
\begin{eqnarray}
G(1,0,0) &=&0.949K_{0}+0.030K  \nonumber \\
G(0,1,0) &=&0.849K_{0}-0.034K \\
G(0,0,1) &=&-0.055K_{0}+1.702K  \nonumber
\end{eqnarray}%
For reference, we also quote the first order results for the equilibrium ($%
E=0$) correlations: 
\begin{eqnarray}
G^{eq}(1,0,0) &=&G^{eq}(0,1,0)=K_{0} \\
G^{eq}(0,0,1) &=&2K  \nonumber
\end{eqnarray}

In the following graphs (Fig. 4a-c), we show the drive dependence of the
three {\em nearest-neighbor} correlation functions, at $K_{0}=1$ and $K=\pm
1 $, to illustrate their behavior in two typical domains (attractive and
repulsive cross-layer coupling). Of course, these values of $K$ and $K_{0}$
are not ``small'', but in a linear approximation they just serve to fix a
scale. Consistent with the interpretation of the drive as an additional
noise which tends to break bonds, all correlations are reduced compared to
their equilibrium value. Further, as the field is switched on, the $%
J\rightarrow -J$ symmetry of the equilibrium system is broken, and the
correlations for repulsive and attractive cross-layer coupling differ from
one another. The details of how they differ provides some insight into the
ordered phases which will eventually emerge.

The first plot (Fig. 4a) shows the correlation function for a
nearest-neighbor bond within a given plane, aligned with the drive
direction. It is interesting to note that the correlations for repulsive
cross-layer coupling are more strongly suppressed than their counterparts
for attractive $J$. This feature becomes more transparent when we consider
nearest-neighbor correlations transverse to the drive, but still within the
same plane (Fig. 4b). For attractive cross-layer coupling, we note that $%
G(1,0,0)$ is considerably enhanced over $G(0,1,0)$, while the two
correlations are roughly equal in the repulsive case. This indicates a
tendency to form droplets of correlated spins which are elongated in the
field direction for $J=+1$ while remaining approximately isotropic for $J=-1$%
, hinting at the nature of the associated ordered phases (strip-like vs
uniform within each layer). This picture is completed when we consider the
cross-plane correlations $G(0,0,1)$ (Fig. 4c): These are positive in the
attractive, and negative in the repulsive case, demonstrating the tendency
towards equal vs opposite local magnetizations on the two layers. Given that
we have performed only a first order calculation, the results really carry a
remarkable amount of information about the system. Encouraged by these
observations, we now consider two other quantities, namely, the particle and
energy currents.

\begin{figure}[tbp]
\input epsf 
\vspace{-1.cm}
\begin{minipage}{0.28\textwidth}
  \epsfxsize = 1.\textwidth \epsfysize = 1.\textwidth \hfill
  \epsfbox{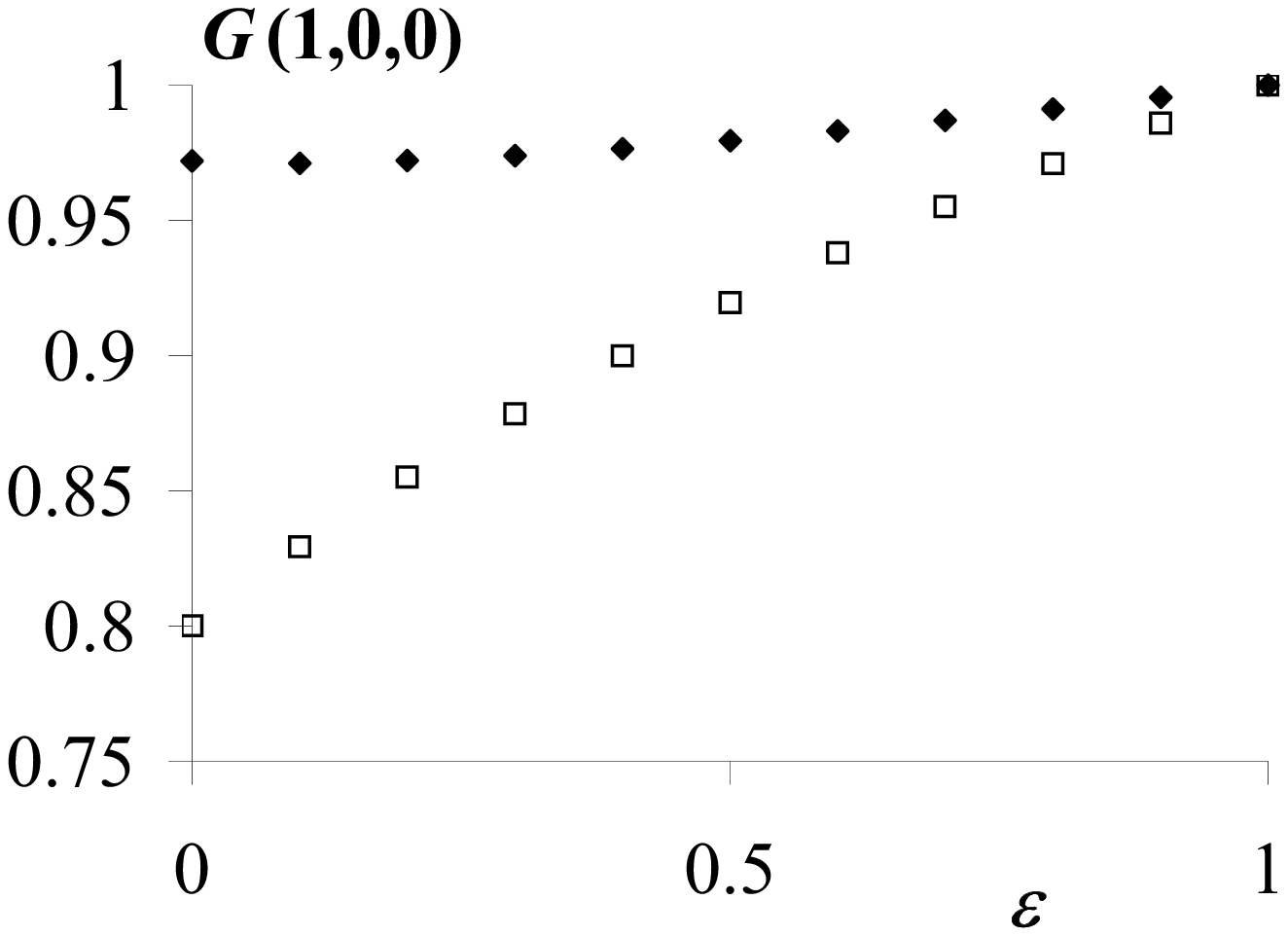} \hfill
    \begin{center} (a) \end{center}
  \end{minipage}
\hfill \hfill 
\begin{minipage}{0.28\textwidth}
  \epsfxsize = 1.\textwidth \epsfysize = 1.\textwidth \hfill
  \epsfbox{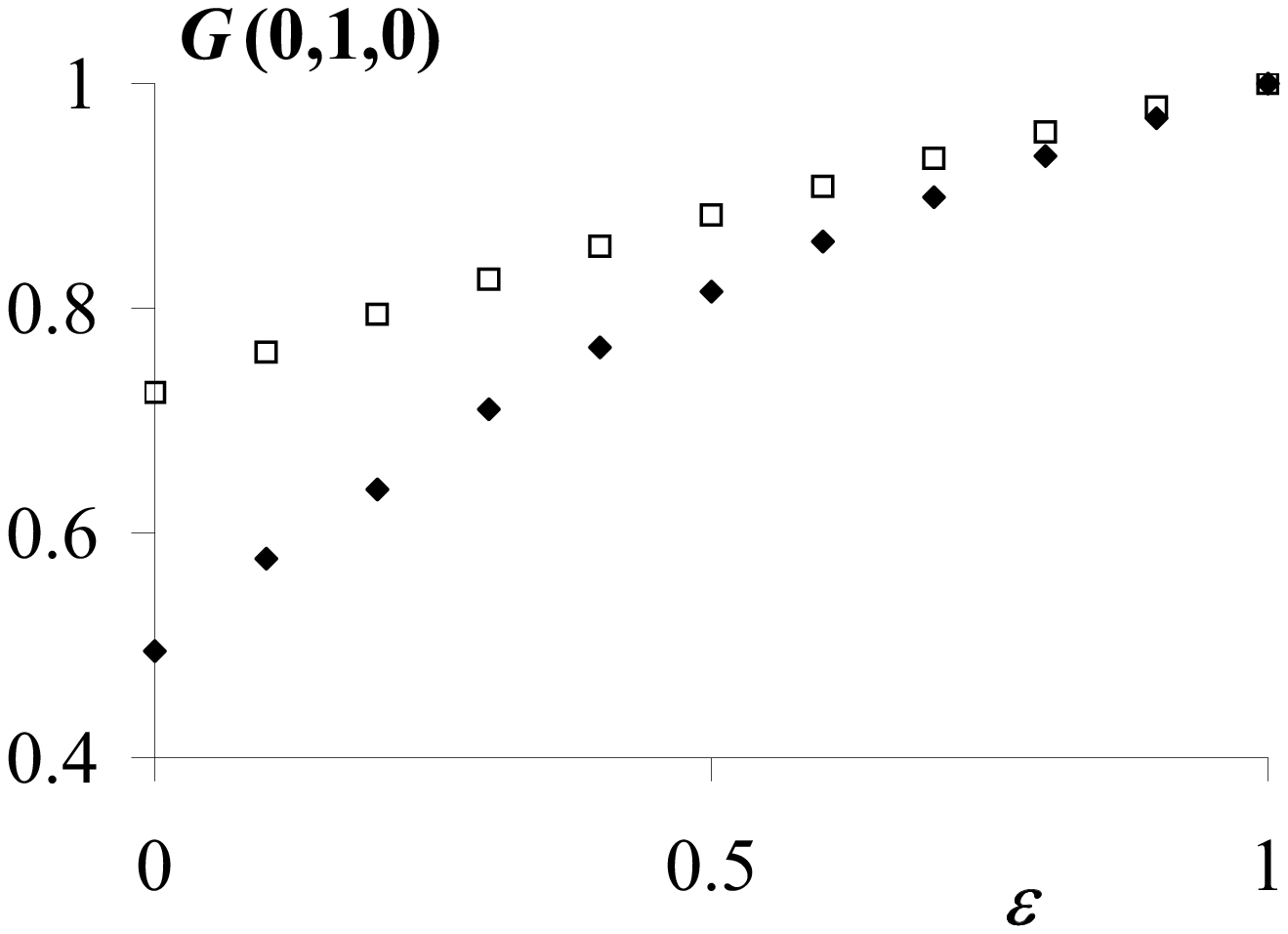} \hfill
    \begin{center} (b) \end{center}
  \end{minipage}
\hfill \hfill \vspace{0.02\textwidth}
\begin{minipage}{0.28\textwidth}
  \epsfxsize = 1.\textwidth \epsfysize = 1.\textwidth \hfill 
  \epsfbox{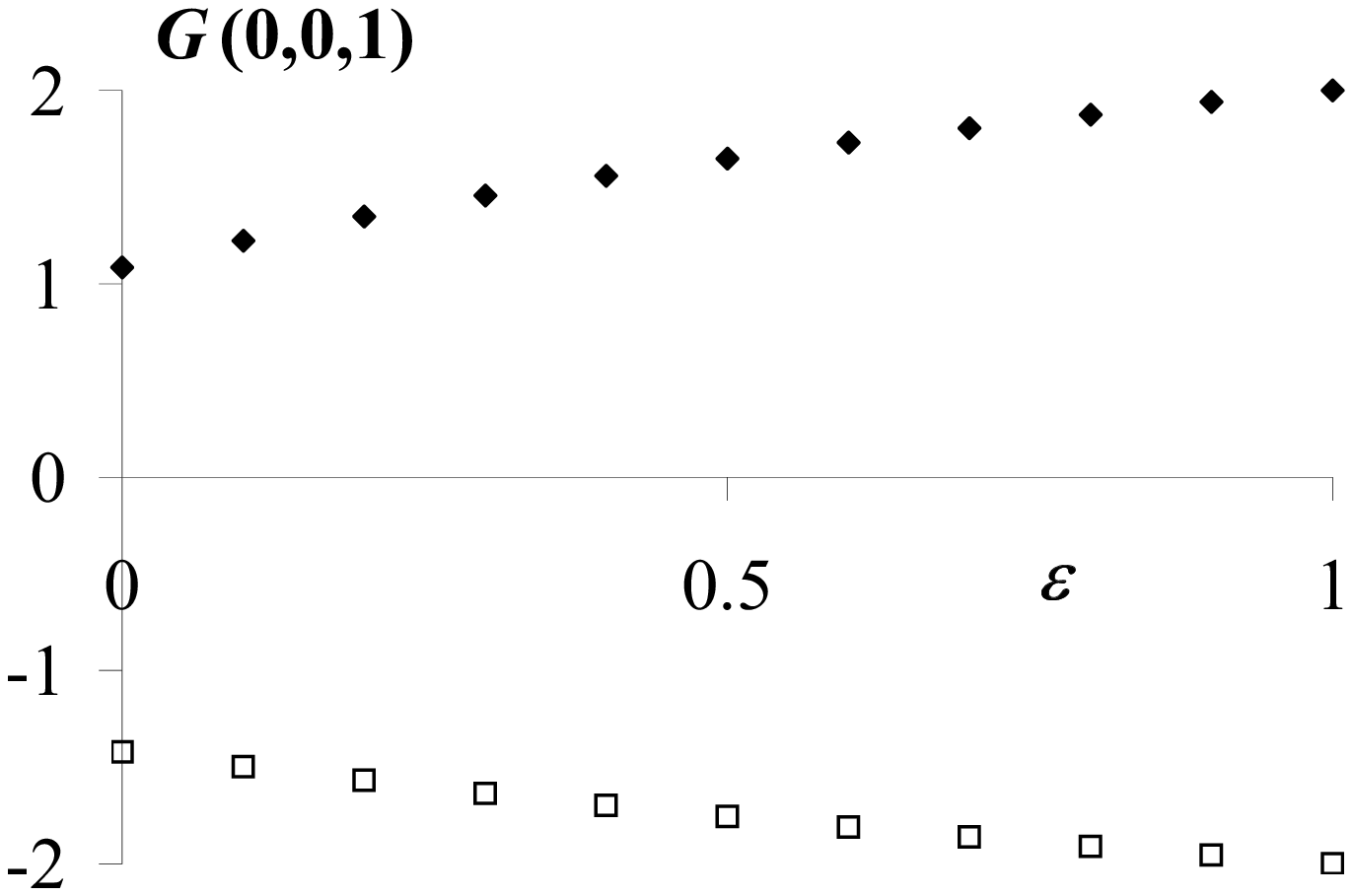}
    \begin{center} (c) \end{center}
  \end{minipage}
\hfill \hfill 
\caption{Nearest-neighbor pair correlations along the $x$- (a), $y$- (b),
and $z$- (c) axes, as functions of the drive parameter $\protect\varepsilon $%
, for $K_{0}=\left| K\right| =1$. Filled diamonds (open squares) refer to
ferromagnetic (antiferromagnetic) cross-layer coupling $K=+1$ ($K=-1$).}
\end{figure}

{\em The particle current. }Due to the bias in conjunction with periodic
boundary conditions, the bilayer system carries a net particle current, $%
\left\langle j\right\rangle $. Since only nearest-neighbor exchanges are
possible, this current is proportional to the density (number per site) of
available particle-hole pairs in the field direction. The transition rate $%
c_{\Vert }$ along this direction, given in Eq.~(\ref{c_par}), then counts
the fraction of these pairs which will actually exchange per unit time.
Specifically, in configuration $\{s\}$, the particle current can be written as
\begin{equation}
j(\{s\})=\frac{1}{2L^{2}}\sum_{\vec{r}}\frac{s(\vec{r})-s(\vec{r}+\hat{x}%
)}{2}c_{\Vert }(\vec{r},\vec{r}+\hat{x};\{s\})\newline
\end{equation}%
For infinite $E$, this expression simplifies considerably, since jumps
against the field will be completely suppressed.

After a few straightforward algebraic manipulations, the {\em average current%
} can be expressed through the pair correlations along the field direction.
To {\em first order} in $K$ and $K_{0}$, we obtain%
\begin{equation}
\left\langle j\right\rangle =%
{\displaystyle{1 \over 4}}%
(1-\varepsilon )[1-G(1,0,0)]+O(K^{2},K_{0}^{2},K_{0}K)
\end{equation}%
which shows that it is non-zero only if $E\neq 0$. Further, it takes its
maximum value at infinite temperature and is reduced by (attractive)
nearest-neighbor interactions. The graph (Fig. 5) shows the field-dependence
of this current, for $K_{0}=1$ and $K=\pm 1$. Since nearest-neighbor
correlations along the field are much larger for positive $J$, indicating a
predominance of particle-particle or hole-hole pairs, the current is reduced
relative to the repulsive case.

\begin{figure}[tbp]
\input{epsf}
\begin{center}
\vspace{-2.5cm}
\hspace{2.5cm}
\begin{minipage}{0.7\textwidth}
  \epsfxsize = 0.7\textwidth \epsfysize = 0.7\textwidth 
  \epsfbox{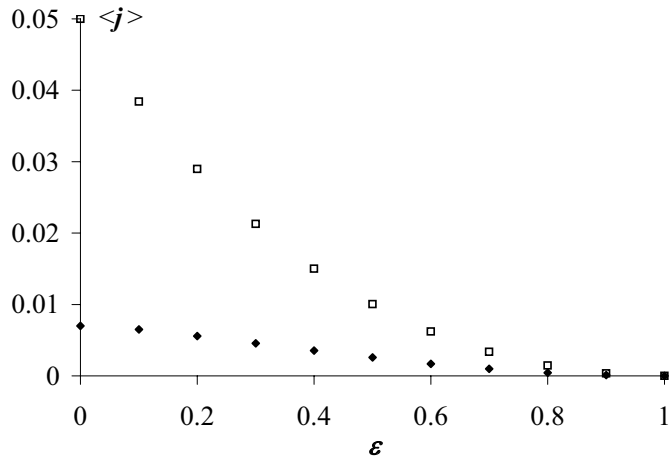}
\end{minipage}
\end{center}
\caption{The average particle current, $\left\langle j\right\rangle $, vs
the drive parameter $\protect\varepsilon $, for $K_{0}=\left| K\right| =1$.
Filled diamonds (open squares) refer to ferromagnetic (antiferromagnetic)
cross-layer coupling.}
\end{figure}

{\em Energy currents. }Another interesting quantity associated with driven
dynamics is the change in configurational {\em energy} during one Monte
Carlo step. In the steady state, by definition, the average configurational
energy is of course constant. However, particle-hole exchanges {\em parallel}
to the field direction tend to increase the energy, since the drive can
easily break bonds. In contrast, exchanges {\em transverse} to $E$ are
purely energetically driven and hence, prefer to satisfy bonds so that the
energy decreases \cite{SZ-rev}. In summary, we have \newline
\begin{equation}
\left\langle 
{\displaystyle{dH \over dt}}%
\right\rangle _{\Vert }=-\left\langle 
{\displaystyle{dH \over dt}}%
\right\rangle _{\perp }>0
\end{equation}%
Even if a particle current were absent, the presence of energy currents
would signal the{\em \ non-equilibrium }steady state.

Since the configurational energy involves only nearest-neighbor bonds, it is
obvious that only the time evolution of nearest-neighbor correlations plays
a role in these two fluxes. Specifically, we have 
\begin{equation}
L^{-2}\left\langle 
{\displaystyle{dH \over dt}}%
\right\rangle _{\Vert }=-J_{0}\newline
\left( \frac{\partial }{\partial t}\right) _{\Vert }\,\left[
G(1,0,0)+G(0,1,0)\right] -2J\left( \frac{\partial }{\partial t}\right)
_{\Vert }\,G(0,0,1)
\end{equation}%
where the subscript on the time derivatives reminds us to select only those
processes which are due to parallel exchanges alone. These can be easily
identified from the terms contributing to Eq.~(\ref{eom}) or (\ref{G-10}).
Of course, there is an analogous equation for $\left\langle
dH/dt\right\rangle _{\perp }$. Collecting the relevant contributions and
multiplying both sides by the inverse temperature $\beta $ to express
everything in terms of $K_{0}$ and $K$, we find:

\begin{eqnarray}
L^{-2}\left\langle 
{\displaystyle{d\beta H \over dt}}%
\right\rangle _{\Vert } &=&-K_{0}\left\{ (1+\varepsilon )\left[
G(2,0,0)-G(1,0,0)\right] +2(1+\varepsilon )\left[ G(1,1,0)-G(0,1,0)\right]
+6\varepsilon K_{0}\right\}  \nonumber \\
&&-2K\left\{ 2(1+\varepsilon )\left[ G(1,0,1)-G(0,0,1)\right] +8\varepsilon
K\right\}
\end{eqnarray}%
The correlation functions spanning next- and next-next nearest
neighbors which appear here can again be determined from our solution for
the structure factors (see Appendix). The result, at $K_{0}=1$ and $K=\pm 1$, 
is shown in Fig.~6
as a function of $\varepsilon $. As expected, this flux is always
non-negative and monotonically increasing as a function of $E$. We note that
the current for $K=-1$ is slightly larger than its counterpart for $K=+1$.
Since it is a complicated function of the couplings and several
correlations, we cannot offer a simple intuitive explanation of this
property.

\begin{figure}[tbp]
\input{epsf}
\begin{center}
\vspace{-2.5cm}
\hspace{2.5cm}
\begin{minipage}{0.7\textwidth}
  \epsfxsize = 0.7\textwidth \epsfysize = 0.7\textwidth 
  \epsfbox{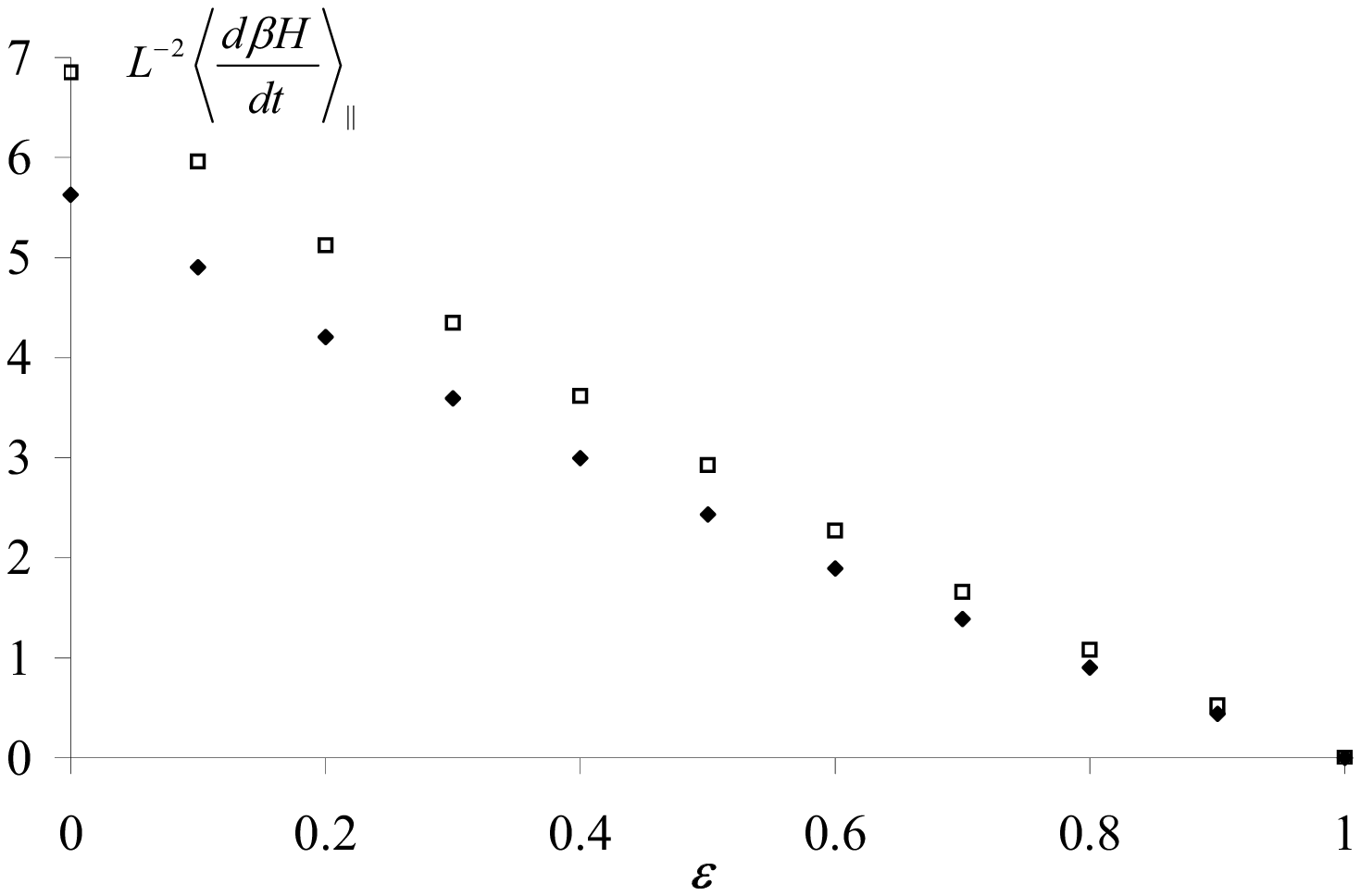}
\end{minipage}
\end{center}
\caption{The average energy current, 
$ L^{-2} \left\langle d\beta H/dt\right\rangle_{\Vert }$, 
vs the drive parameter $\protect\varepsilon $, for $K_{0}=\left|
K\right| =1$. Filled diamonds (open squares) refer to ferromagnetic
(antiferromagnetic) cross-layer coupling.}
\end{figure}

\section{Concluding remarks}

Based directly on the microscopic lattice dynamics, the high temperature
series provides us with a simple analytic tool which complements field
theoretic approaches. Even in a first order approximation, it is remarkable
how many features of the MC results are -- at least qualitatively --
reproduced. To summarize our results briefly, we derive, and solve, a set of
equations for the stationary pair correlation functions and their Fourier
transforms, the equal-time structure factors. By matching the series
expansion of the latter with the expected critical singularity, we find two
critical lines, separating the disordered phase from the strip phase (S) and
the full-empty phase (FE), respectively. We also observe the shift of the
bicritical point which marks the juncture of these two lines, in very good
qualitative agreement with the simulations. To illustrate the
non-equilibrium character of the steady state, we compute the particle
current and the energy flux through the system. The particle current is
determined by the nearest-neighbor correlations in the field direction, and
takes its maximum value in the absence of interactions. Our findings for the
energy current confirm intuitive expectations: parallel exchanges tend to
lower, while transverse exchanges tend to increase, the configurational
energy.

A brief comment on boundary conditions is in order. Even though it is quite
natural to use periodic boundary conditions in all lattice directions, it is
not unreasonable to consider other choices, especially in the $z$-direction.
To recall, periodicity in $z$ implies that the site ($x,y,0$) is connected
to the site ($x,y,1$) via {\em two} bonds which enter into {\em both} the
energetics {\em and} and the dynamics (i.e., there are two channels for a
particle to move from one layer to the other). Alternately, we can choose
open boundary conditions in $z$ and consider only a single energetic bond
and single dynamical channel between these two sites. Mixtures of these two
cases can also be constructed: i.e., imposing periodic boundary conditions
on the energetics, but allowing only a single channel for particle moves, or
vice versa. The first (second) ``mixed'' case is reducible to the case of
open (periodic) boundary conditions, with $J$ replaced by $2J$ ($J/2$). Even
though details are not presented here, we did, in fact, compute the critical
lines for different cross-plane boundary conditions. The main conclusions of
our study, namely, the existence of the two continuous phase transitions and
the shift of the bicritical point, hold for all of these variations.

The high temperature expansion presented here has two shortcomings. First,
our results provide no insight into the first-order transitions between the
FE and S phases which were observed in the simulations. As in all high
temperature series, the first singularity which is encountered as $T$ is
lowered determines the radius of convergence. A low-temperature approach
would be necessary to capture transitions between ordered phases. Second,
our series is currently limited to just one nontrivial term. In order to
compute the second order correction to the pair correlations, we would need
to know the full stationary distribution, $P^{\ast }$, to first order.
Writing the stationary master equation in the form $0=LP^{\ast }$ where $L$
is the linear operator (``Liouvillean'') defined by the transition rates,
this requires the full inverse of $L$, to zeroth order. Finding this inverse
is a highly nontrivial (and as yet unsolved) problem.

Inspite of these drawbacks, the high temperature expansion is one of the few
analytic tools which provide insight into non-equilibrium steady states. It
is conceptually and mathematically straightforward, and -- at least at the
qualitative level -- surprisingly reliable. Since it is based directly on
the microscopic lattice dynamics, it still carries information about
nonuniversal properties which would be lost upon taking a continuum limit.
It is therefore a valuable complement to both simulations and field
theoretic methods.

{\bf Acknowledgements.} We thank U.C. T\"{a}uber and R.K.P. Zia for fruitful
discussions. Financial support from the NSF through the Division of
Materials Research is gratefully acknowledged.

\newcommand{\appendixname}{Appendix}
\appendix
\section{}
\vspace{-0.2cm}
\subsection{Equations for the two-point correlations.}
\vspace{-0.2cm}

To illustrate the general procedure, we provide a few details here \cite%
{ZWLV}. As an example, we choose the two-point correlation $G(1,1,0)$. We
start with the equation of motion for the pair correlations:

\[
{\displaystyle{d \over dt}}%
\left\langle s(\vec{r})s(\vec{r}\,^{\prime })\right\rangle =\sum_{\vec{x},%
\vec{x}^{\prime }}\left\langle s(\vec{r})s(\vec{r}\,^{\prime })\left[ s(\vec{%
x})s(\vec{x}\,^{\prime })-1\right] \,c\left( \vec{x},\vec{x}\,^{\prime
};\left\{ s\right\} \right) \right\rangle 
\]%
where the sum runs over nearest-neighbor pairs ($\vec{x},\vec{x}\,^{\prime }$%
) such that $\vec{x}\in \{\vec{r},\vec{r}\,^{\prime }\}$ but $\vec{x}%
\,^{\prime }\notin \{\vec{r},\vec{r}\,^{\prime }\}$. To obtain the equation
satisfied by $G(1,1,0)$, we choose, e.g. $\vec{r}\equiv (0,0,0)$ and $\vec{r}%
\,^{\prime }\equiv (1,1,0)$. The two participating spins have a total of $8$
distinct nearest neighbors: $6$ of these lie in the $z=0$ plane, and $2$ are
found on the $z=1$ plane. One possible ($\vec{x},\vec{x}\,^{\prime }$) pair
is, for example, the pair $\vec{x}\equiv \vec{r}$ and $\vec{x}\,^{\prime
}\equiv (1,0,0)$. The corresponding exchange occurs along the field
direction, and hence, we must use the rate $c_{\Vert }$ from Eq.~(\ref{c_par}%
). Considering {\em only} the contribution due to this particular pair of
sites, we obtain 
\begin{eqnarray*}
\partial _{t}G(1,1,0) &=&%
{\displaystyle{1 \over 4}}%
\left\langle s(0,0,0)s(1,1,0)\left[ s(0,0,0)s(1,0,0)-1\right] \right. \\
&&\times \left. \left\{ \left[ s(0,0,0)-s(1,0,0)+2\right] +\varepsilon \left[
s(1,0,0)-s(0,0,0)+2\right] \exp (-\beta \Delta H)\right\} \right\rangle +...
\end{eqnarray*}%
Here, $\ ...$ stands for the contributions due to all other possible ($\vec{x%
},\vec{x}\,^{\prime }$) pairs which can be handled in an analogous manner.
After multiplying out a few terms and neglecting $3$-point functions, the
expression above simplifies to 
\begin{eqnarray*}
\partial _{t}G(1,1,0) &=&%
{\displaystyle{1 \over 2}}%
\left[ G(0,1,0)-G(1,1,0)\right] \\
&&+%
{\displaystyle{\varepsilon  \over 4}}%
\left\langle \left[ s(1,0,0)s(1,1,0)-s(0,0,0)s(1,1,0)\right] \left[
s(1,0,0)-s(0,0,0)+2\right] \exp (-\beta \Delta H)\right\rangle +...
\end{eqnarray*}%
Note that we have replaced $\left\langle s(1,0,0)s(1,1,0)\right\rangle $ by
the corresponding correlation function, $G(0,1,0)$. Next, we expand $\exp
(-\beta \Delta H)$ in powers of $K$ and $K_{0}$, according to Eq.~(\ref%
{c_par_exp}). The zeroth order contribution is easily accounted for, leaving
us with the $O(\beta )$ correction: 
\begin{eqnarray*}
\partial _{t}G(1,1,0) &=&%
{\displaystyle{1 \over 2}}%
\left( 1+\varepsilon \right) \left[ G(0,1,0)-G(1,1,0)\right] \\
&&-\beta 
{\displaystyle{\varepsilon  \over 4}}%
\left\langle \left[ s(1,0,0)s(1,1,0)-s(0,0,0)s(1,1,0)\right] \left[
s(1,0,0)-s(0,0,0)+2\right] \left( \Delta H\right) \right\rangle +...
\end{eqnarray*}%
The change in energy, $\Delta H$, involves the nearest-neighbor spins of the
selected pair. Since these terms are already explicitly first order in $%
\beta $, they are averaged using the zeroth order approximation to the
steady state solution. The latter is uniform so that the averages are
trivial. Collecting, we find that the contribution of this particular
exchange to $G(1,1,0)$ is 
\[
\partial _{t}G(1,1,0)=%
{\displaystyle{1 \over 2}}%
\left( 1+\varepsilon \right) \left[ G(0,1,0)-G(1,1,0)\right] -\varepsilon
K_{0}+... 
\]%
Including all the other ($\vec{x},\vec{x}\,^{\prime }$) pairs, we arrive at
the complete equation: 
\begin{eqnarray}
\partial _{t}G(1,1,0) &=&(1+\varepsilon
)[G(2,1,0)+G(0,1,0)-2G(1,1,0)]+2[G(1,2,0)+G(1,0,0)  \nonumber \\
&&-2G(1,1,0)]+\newline
4[G(1,1,1)-G(1,1,0)]-2K_{0}-2\varepsilon K_{0}\newline
\nonumber
\end{eqnarray}%
Of course, this procedure is easily extended to any other two-point
function. Moreover, it is straightforward to track which terms in the
equations arise from parallel, and which from transverse, exchanges. This
distinction is crucial for the computation of the energy fluxes.
\vspace{-0.2cm}

\subsection{Matrix inversion.}
\vspace{-0.2cm}
\ We seek to invert the system of equations (\ref{matrix}) for the three
unknown expressions $I_{1}$, $I_{2}$, and $I_{3}$. We follow the method
first outlined in \cite{SZ}. With a little algebra, it becomes apparent that
the coefficients and inhomogeneities in these equations involve integrals of
the form 
\begin{eqnarray*}
Q_{lmn}(\varepsilon )\equiv \int \frac{(1-\cos k)^{l}(1-\cos p)^{m}(1-\cos
q)^{n}}{\delta }  \nonumber
\end{eqnarray*}
with non-negative integer $l,m,n$ and $l+m+n\leq 3$. The task of determining
these integrals is simplified by a series of relations, namely, 
\begin{eqnarray*}
1 &=&\int 
{\displaystyle{\delta  \over \delta }}%
=2(1+\varepsilon )Q_{100}+4Q_{010}+4Q_{001} \\
1 &=&\int (1-\cos k)%
{\displaystyle{\delta  \over \delta }}%
=2(1+\varepsilon )Q_{200}+4Q_{110}+4Q_{101} \\
1 &=&\int (1-\cos p)%
{\displaystyle{\delta  \over \delta }}%
=2(1+\varepsilon )Q_{110}+4Q_{020}+4Q_{011} \\
{{\frac{3}{2}}} &=&\int (1-\cos k)^{2}%
{\displaystyle{\delta  \over \delta }}%
=2(1+\varepsilon )Q_{300}+4Q_{210}+4Q_{201} \\
{{\frac{3}{2}}} &=&\int (1-\cos p)^{2}%
{\displaystyle{\delta  \over \delta }}%
=2(1+\varepsilon )Q_{120}+4Q_{030}+4Q_{021} \\
1 &=&\int (1-\cos k)(1-\cos p)%
{\displaystyle{\delta  \over \delta }}%
=2(1+\varepsilon )Q_{210}+4Q_{120}+4Q_{111} \\
1 &=&\int (1-\cos q)%
{\displaystyle{\delta  \over \delta }}%
=2(1+\varepsilon )Q_{101}+4Q_{011}+4Q_{002}
\end{eqnarray*}%
The computation of the remaining integrals, while tedious, is completely
straightforward. Once these coefficients are known, Eqns (\ref{matrix}) can
be inverted. 
\vspace{-0.2cm}

\subsection{The equilibrium solution}
\vspace{-0.2cm}

Since our expansion presumed $%
E>\Delta H$, we may not simply set $\varepsilon =1$ in our equations of
motion for the two-point correlations. Instead, one should rederive the
whole set carefully, noting the absence of the driving field. Of course,
this is trivial, since only the first order term in the expansion of the
equilibrium (Boltzmann) distribution is required here. In this order, only
nearest-neighbor correlations can be nonzero, so that the only non-vanishing 
$G$'s are 
\begin{eqnarray*}
G^{eq}(0,0,0) &=&1 \\
G^{eq}(\pm 1,0,0) &=&G^{eq}(0,\pm 1,0)=K_{0} \\
G^{eq}(0,0,1) &=&2K
\end{eqnarray*}%
Performing the Fourier transform to structure factors and exploiting the
boundary conditions in the $z$-direction, we find immediately that 
\begin{eqnarray*}
S(k,p,q) &=&\sum_{z=0,1}\sum_{x,y=-\infty }^{\infty }G(x,y,z)e^{-i(kx+py+qz)}
\\
&=&1+2K_{0}\left( \cos k+\cos p\right) +2K\cos q
\end{eqnarray*}%
resulting in 
\begin{eqnarray*}
\lim_{k,p\rightarrow 0}S^{-1}(k,p,q) &=&1-2K_{0}\left( 2-\frac{1}{2}k^{2}-%
\frac{1}{2}p^{2}\right) -2K\cos q+O(k^{4},p^{4}) \\
&\equiv &\tau (q)+O(k^{2},p^{2})
\end{eqnarray*}%
with 
\[
\tau (q)=1-4K_{0}-2K\cos q 
\]%
For $q=0$, this vanishes at 
\[
\ k_{B}T_{c}^{S}=4J_{0}+2J 
\]%
and for $q=\pi $, the zero shifts to 
\[
\ k_{B}T_{c}^{FE}=4J_{0}-2J 
\]%
Thus, D-S transitions are observed for $J>0$, and D-FE transitions dominate
for $J<0$. The bicritical point is located at $k_{B}T_{c}(J=0)=4J_{0}$.
\vspace{-0.2cm}

\subsection{Other correlations near the origin.}
\vspace{-0.2cm}

These are required to compute the energy fluxes along the parallel and
transverse directions. \ Specifically, we need the following correlation
functions: $G(1,1,0)$, $G(1,0,1)$, $G(0,1,1)$, $G(2,0,0)$, and $G(0,2,0)$.
We want to write these correlation functions in terms of the already
calculated integrals $I_{1},I_{2},I_{3}$ and also in terms of the set of $%
Q_{lmn}$ integrals defined earlier.

We start with the definition for $G(2,0,0)$ and substitute the expression
for the structure factor: 
\begin{eqnarray*}
\ G(2,0,0) &=&\int \tilde{S}\exp (2ik)=2\int \tilde{S}(1-\cos k)^{2}-4I_{1}%
\newline
\\
&=&2\left\{ 2(1+\varepsilon )I_{1}Q_{200}+4\varepsilon
(5K_{0}+2K)Q_{300} \right. \\
&& + \left. 4(I_{2}+5K_{0}+2K)Q_{210}+4\left( I_{3}+4K_{0}\right)
Q_{201} \right. \\
&& - \left. 8K_{0}(1+\varepsilon )Q_{310}-8\left( K+K_{0}\right) Q_{211} \right.
\\
&& - \left. 8\left( \varepsilon K+K_{0}\right) Q_{301}-8\varepsilon
K_{0}Q_{400}-8K_{0}Q_{220}-2I_{1}\right\}
\end{eqnarray*}%
and similarly, 
\begin{eqnarray*}
G(0,2,0) &=&2\int \tilde{S}(1-\cos p)^{2}-4I_{2}\newline
\\
&=&2\left\{ 2(1+\varepsilon )I_{1}Q_{020}+4\varepsilon
(5K_{0}+2K)Q_{120} \right. \\
&& + \left. 4(I_{2}+5K_{0}+2K)Q_{030}+4\left( I_{3}+4K_{0}\right)
Q_{021} \right. \\
&& - \left. 8K_{0}(1+\varepsilon )Q_{130}-8\left( K+K_{0}\right) Q_{031} \right.
\\
&& - \left. 8\left( \varepsilon K+K_{0}\right) Q_{121}-8\varepsilon
K_{0}Q_{220}-8K_{0}Q_{040}-2I_{2}\right\}
\end{eqnarray*}%
The remaining correlation functions follow in the same way: 
\begin{eqnarray*}
G(1,1,0) &=&\int \tilde{S}(1-\cos k)(1-\cos p)-I_{1}-I_{2}\newline
\  \\
&=&\left\{ 2(1+\varepsilon )I_{1}Q_{110}+4\varepsilon
(5K_{0}+2K)Q_{210}\right. \\
&&\left. +4(I_{2}+5K_{0}+2K)Q_{120}+4\left( I_{3}+4K_{0}\right)
Q_{111}\right. \\
&&\left. -8K_{0}(1+\varepsilon )Q_{220}-8\left( K+K_{0}\right) Q_{121}\right.
\\
&&\left. -8\left( \varepsilon K+K_{0}\right) Q_{211}-8\varepsilon
K_{0}Q_{310}-8K_{0}Q_{130}-I_{1}-I_{2}\right\}
\end{eqnarray*}%
\begin{eqnarray*}
G(1,0,1) &=&\int \tilde{S}(1-\cos k)(1-\cos q)-I_{1}-I_{3}\newline
\  \\
&=&\left\{ 2(1+\varepsilon )I_{1}Q_{101}+4\varepsilon
(5K_{0}+2K)Q_{201}\right. \\
&&\left. +4(I_{2}+5K_{0}+2K)Q_{111}+4\left( I_{3}+4K_{0}\right)
Q_{102}\right. \\
&&\left. -8K_{0}(1+\varepsilon )Q_{211}-8\left( K+K_{0}\right) Q_{112}\right.
\\
&&\left. -8\left( \varepsilon K+K_{0}\right) Q_{202}-8\varepsilon
K_{0}Q_{301}\right. \\
&&\left. -8K_{0}Q_{121}-I_{1}-I_{3}\right\}
\end{eqnarray*}%
\begin{eqnarray*}
\ G(0,1,1) &=&\int \tilde{S}(1-\cos p)(1-\cos q)-I_{2}-I_{3} \\
&=&\left\{ 2(1+\varepsilon )I_{1}Q_{011}+4\varepsilon
(5K_{0}+2K)Q_{111}\right. \\
&&\left. +4(I_{2}+5K_{0}+2K)Q_{021}+4\left( I_{3}+4K_{0}\right)
Q_{012}\right. \\
&&\left. -8K_{0}(1+\varepsilon )Q_{121}-8\left( K+K_{0}\right) Q_{022}\right.
\\
&&\left. -8\left( \varepsilon K+K_{0}\right) Q_{112}-8\varepsilon
K_{0}Q_{211}\right. \\
&&\left. -8K_{0}Q_{031}-I_{2}-I_{3}\right\}
\end{eqnarray*}%
After the additional $Q$-integrals have been determined, the energy currents
are easily found.


\vspace{-0.5cm}

\begin{references}
\vspace{-1.2cm}
\bibitem{SZ-rev} B. Schmittmann and R.K.P Zia, in {\em Phase Transitions and
Critical Phenomena}, \ Vol 17, eds. C. Domb and J.L. Lebowitz (Academic
Press, London, 1995).

\bibitem{other-revs} D. Mukamel, in {\em Soft and Fragile Matter:\
Nonequilibrium Dynamics, Metastability and Flow}, eds. M.E. Cates and M.R.
Evans (Institute of Physics Publishing, Bristol, 2000); J. Marro and R.
Dickman, {\em Nonequilibrium Phase Transitions in Lattice Models} (Cambridge
University Press, Cambridge, 1999).

\bibitem{KLS} S. Katz, J.L. Lebowitz, and H. Spohn, {\em Phys. Rev. B} {\bf %
28}:1655 (1983) and and{\em \ J. Stat. Phys.} {\bf 34}:497 (1984).

\bibitem{ZWLV} M.Q. Zhang, J.-S. Wang, J.-L. Lebowitz, and J.L. Vall\`{e}s, 
{\em J. Stat. Phys.} {\bf 52}:1461 (1988). 

\bibitem{GLMS} P.L. Garrido, J.L. Lebowitz, C. Maes, and H. Spohn, 
{\em Phys. Rev. A} {\bf 42}:1954 (1990).

\bibitem{crit} H.-K. Janssen and B. Schmittmann, {\em Z. Phys. B} {\bf 64}%
:503 (1986); K.-t. Leung and J.L. Cardy, {\em J. Stat. Phys.} {\bf 44}:567
(1986) and {\bf 45}:1087 (1986) (erratum).

\bibitem{lowT} J.L. Vall\`{e}s, K.-t. Leung, and R.K.P. Zia, {\em J. Stat.
Phys.} {\bf 56}:43 (1989).

\bibitem{KKM} K.K. Mon, private communication (1991).

\bibitem{2l-early} A. Achahbar, P.L Garrido and J. Marro, {\em Phys Lett. A }%
{\bf 172}:29 (1992); A. Achahbar and J. Marro,{\it \ }{\em J. \ Stat. \ Phys.%
} {\bf 78}:1493 (1995).

\bibitem{HZS} C.C. Hill, R.K.P. Zia and B.Schmittmann, {\it Phys. Rev. Lett. 
}{\bf 77}:514 (1996). See also B. Schmittmann, C.C. Hill, and R.K.P. Zia, 
{\em Physica A} {\bf 239}:382 (1997).

\bibitem{CW} C.-P. Chng and J.-S. Wang, {\em Phys. Rev. E} {\bf 61}:4962
(2000).

\bibitem{TSZ} U.C. T\"{a}uber, B. Schmittmann, and R.K.P. Zia {\em J. Phys. A%
}{\bf \ 34:}L583 \ (2001).

\bibitem{ballentine} L. E. Ballentine {\em Physica }{\bf 30:}1231(1964).

\bibitem{binder} K. Binder {\em Thin Solid Films}{\it \ }{\bf 20:}367(1974).

\bibitem{hansen} P.L. Hansen, J. Lemmich, J.H. Ipsen, and O.G. Mouritsen, 
{\em J. Stat. Phys.} {\bf 73}:723 (1993). This article also gives a brief
history and further references.

\bibitem{dim-cross} T.W. Capehart and M.E. Fisher, {\em Phys. Rev. B} {\bf 13%
}:5021 (1976).

\bibitem{intercalation} M. S. Dresselhaus and G. Dresselhaus, {\em Adv. Phys.%
} {\bf 30}:139 (1981); G. R. Carlow and R. F. Frindt, {\em Phys. Rev. B} 
{\bf 50}:11107 (1994). See also G. R. Carlow, {\em Intercalation Channels in
Staged Ag Intercalated TiS}$_{2}.$ Ph.D Thesis, Simon Frasier University
(1992).

\bibitem{ferrenberg} A. Ferrenberg and D.P. Landau, {\em J. Appl. Phys. } 
{\bf 70}:6215 (1991).

\bibitem{HTS-revs} See, especially, C. Domb, and D.S. Gaunt and A.J.
Guttmann, in {\em Phase Transitions and Critical Phenomena}, \ Vol 3, eds C.
Domb and M.S.\ Green (Academic, London, 1974); and A.J.Guttmann, in {\em %
Phase Transitions and Critical Phenomena}, \ Vol 13, eds. C. Domb and J.L.
Lebowitz (Academic Press, London, 1989).

\bibitem{SZ} B. Schmittmann and R.K.P. Zia, {\em J. Stat.Phys.} {\bf 91}%
:525(1998).

\bibitem{LZS} L.B. Shaw, B. Schmittmann and R. K. P. Zia, {\em J. Stat.
Phys. }{\bf 95}:981 (1999).

\bibitem{Metropolis} N. Metropolis, A.W. Rosenbluth, M.M. Rosenbluth, A.H.
Teller and E. Teller, {\em J. Chem. Phys.} {\bf 21}:1097 (1953).

\bibitem{Onsager} L. Onsager {\em Phys. Rev.} {\bf 65}:117 (1944); B.M.
McCoy and T.T. Wu, {\em The Two-dimensional Ising Model} (Harvard University
Press, Cambridge, MA, 1973).

\bibitem{spitzer} F. Spitzer, {\em Adv. Math. }{\bf 5}:246 (1970).

\bibitem{AGMA} J.J. Alonso, P.L. Garrido, J. Marro, and A. Achahbar, {\em J.
Phys. A} {\bf 28}:4669 (1995).

\bibitem{CH} C.C. Hill, {\em Phase Transitions in Driven Bi-layer Systems}.
Honors Thesis, Virginia Tech (1996).
\end{references}
\end{document}